# UNIFIED DESCRIPTION OF ACCRETION FLOWS AROUND BLACK HOLES

Xingming Chen[1], Marek A. Abramowicz[1],
Jean-Pierre Lasota[2],
Ramesh Narayan[3], and Insu Yi[3]

[1]Department of Astronomy and Astrophysics, Göteborg University
and Chalmers University of Technology, 412 96 Göteborg, Sweden

[2]UPR 176 du CNRS; DARC, Observatoire de Paris, Section de Meudon,
92195 Meudon Cedex, France

[3]Harvard-Smithsonian Center for Astrophysics, 60 Garden Street, Cambridge, MA 02138

## ABSTRACT

We provide a unified description of thermal equilibria of black hole accretion disks, including the newly-discovered advection-dominated solutions. We classify the solutions on the basis of optical depth and importance of advection cooling. We demonstrate that only four physically distinct topological types of equilibria exist. Two of the types correspond to optically thin and optically thick equilibria, while the other two types are distinguished by whether advection is negligible or dominant. A stable Shakura-Sunyaev disk exists only for accretion rates $\dot{M}$ below a certain maximum. However, there is a critical viscosity parameter $\alpha_{\rm crit}$, which is a function of radius, such that for $\alpha > \alpha_{\rm crit}$ advection-dominated solutions exist for all $\dot{M}$. Even when $\alpha < \alpha_{\rm crit}$, the advection-dominated solutions are available for a wide range of $\dot{M}$ except for a gap around the Eddington rate. We therefore suggest that advection-dominated flows may be more common than standard thin disks in black hole systems. For certain ranges of radii and $\dot{M}$, no stable steady state solution is possible. In these cases, we suggest that limit cycle behavior may occur, leading to variability.

*Subject headings:* accretion, accretion disks — instabilities

# 1. INTRODUCTION

Accretion disks are common in a variety of astrophysical systems, including young stars, cataclysmic variables, X-ray binaries, and active galactic nuclei. A simple model of an accretion disk was developed by Shakura & Sunyaev (1973) based on the assumption that the energy released through viscous dissipation is radiated locally and efficiently, so that the disk vertical thickness is much smaller than the radius. This "thin disk" model has been the mainstay of accretion studies for many years. However, the model is known to suffer from thermal and viscous instabilities under certain circumstances (Pringle et al. 1973, Lightman & Eardley 1974). Shapiro, Lightman & Eardley (1976) suggested that these instabilities sometimes result in a two-temperature disk.

Starting with the initial work of Abramowicz et al. (1988) and the more recent papers of Narayan & Popham (1993), Narayan & Yi (1994, 1995a) and Abramowicz et al. (1995), it has been demonstrated that an unstable accretion disk may switch to a completely different mode of accretion where the viscously released energy is advected with the gas rather than being radiated. The resulting "advection-dominated" flow is stable (Abramowicz et al. 1995, Narayan & Yi 1995b) both to the thermal and viscous instabilities. Advection-dominated flows are much hotter than cooling-dominated thin disks and are therefore thicker in the vertical direction. For this reason they are also described as "thick disks" or "slim disks," depending on the relative vertical thickness.

Since we now have at least two distinct stable accretion solutions, namely cooling-dominated flows and advection-dominated flows, it is important to map out which kind of solution occurs under different conditions. This is the purpose of the present *Letter*. We demonstrate the topological relationships among the various solution branches in the case of accretion onto a black hole, and unify previous discussions by including the new advection-dominated solutions.

# 2. BASIC EQUATIONS

We use the slim disk formalism described in Abramowicz et al. (1995). This model makes the following assumptions: (i) The vertical half-thickness of the disk $H$ is smaller than the radius $R$, i.e. $H(R) = c_s/\Omega_K \lesssim R$, where $c_s = \sqrt{p/\rho}$ is the local sound speed ($p$ is the total pressure and $\rho$ is the density) and $\Omega_K(R)$ is the Keplerian angular velocity calculated by using a pseudo-Newtonian potential (Paczyński & Wiita 1980). (ii) The kinematic viscosity coefficient is written as $\nu = (2/3)\alpha c_s H$. (iii) The disk is assumed to rotate at $\Omega = \Omega_K$.

We use the following three steady state equations:

$$\nu\Sigma = \frac{\dot{M}}{3\pi}f_*g^{-1}, \qquad (angular\ momentum\ conservation) \qquad (2.1)$$

where $\Sigma = 2H\rho$ is the surface density, $f_* = 1 - 9\Omega_*/(\Omega(R/R_G)^2)$ with $R_G = 2GM/c^2$, $g = -(2/3)(d\ln\Omega/d\ln R)$, and $\Omega_*$ is equal to $\Omega_K$ at $R = 3R_G$;

$$\dot{M} = -2\pi R\Sigma v_r, \qquad (mass\ conservation) \qquad (2.2)$$

where $v_r$ is the radial velocity; and

$$Q_+ = Q_- + Q_{\text{adv}}, \qquad (energy\ conservation) \qquad (2.3)$$

where the left hand side is the viscous heating per unit area and the right hand side is the sum of radiative cooling and advective cooling. In conventional cooling-dominated thin disks $Q_{\text{adv}}$ is negligible, but this term dominates in advection-dominated flows. We take

$$Q_+ = \frac{3\dot{M}}{4\pi}\Omega^2 f_*g, \qquad Q_{\text{adv}} = -\Sigma v_r\frac{p}{\rho}\xi = \frac{\dot{M}}{2\pi R^2}\frac{p}{\rho}\xi, \qquad (2.4)$$

where $\xi = -[(4-3\beta)/(\Gamma_3-1)](d\ln T/d\ln R) + (4-3\beta)(d\ln\Sigma/d\ln R)$, $T$ is the disk midplane temperature, $\beta = p_{\text{gas}}/p$ is the ratio of gas to total pressure, $\Gamma_3 = 1 + (4-3\beta)(\gamma-1)/[\beta + 12(\gamma-1)(\beta-1)]$, and $\gamma$ is the ratio of specific heats (Chen & Taam 1993).

For the local radiative cooling, we go beyond the model of Abramowicz et al. (1995) and adopt the bridging formula introduced by Wandel & Liang (1991, cf. Luo & Liang 1994). Their formula takes into account free-free and electron scattering opacities, with Comptonization, and provides an *ad hoc* description of the radiative transfer when the effective optical depth $\tau_{\text{eff}} \approx 1$. A correct description of the $\tau_{\text{eff}} \approx 1$ regime requires solving the problem of the vertical disk structure. This is notoriously difficult and the results depend on the unknown vertical stratification of the viscous heating (e.g. Shaviv and Wehrse 1989). Other phenomenological prescriptions for $Q_-$ at $\tau_{\text{eff}} \approx 1$ (e.g. Lasota & Pelat 1991, Narayan & Yi 1995b, Artemova et al. 1994) do not modify the solution topology, as we show below. Finally, the equation of state is written as the sum of gas and radiation pressure, with $p_{\text{rad}} = Q_-(\tau_R/2c)$, where $\tau_R$ is the Rosseland mean optical depth.

While most of the results presented here make use of the above equations, we also show for comparison some calculations with a different formalism developed by Narayan & Yi (1995b). The major differences in the latter model are (i) the explicit use of a self-similar solution with significant sub-Keplerian rotation in the advection-dominated regime (Narayan & Yi 1994), (ii) a Newtonian rather than the more accurate pseudo-Newtonian potential, (iii) a different vertical integration for the advection term, $Q_{\text{adv}}$, (iv) the inclusion of magnetic pressure in the equation of state, thermal synchrotron cooling and more elaborate Comptonization, (v) allowing the plasma to be two-temperature, and (vi) a different form of $Q_-$ in the regime $\tau_{\rm eff} \approx 1$.

Most of these differences lead to only modest changes. However, the inclusion of Comptonized synchrotron cooling (item iv), which is a very efficient process at high temperatures, has a significant effect especially for $\tau_{\rm eff} \lesssim 1$. Nevertheless, as discussed below, the qualitative form of the results is unaffected compared to the simpler model described earlier. The difference (iii) arises primarily because of different assumptions on the profile of $H(R)$. The form of $\xi$ given in the line below equation (2.4) corresponds to a constant $H$. In the more general case, $\xi$ has an additional term $-(4 - 3\beta)d\ln H/d\ln R$; setting $d\ln H/d\ln R = 1$ recovers the self-similar advection term used in Narayan & Yi (1994). Pure gas flows with $\gamma = 5/3$ give non-rotating self-similar behavior according to Narayan & Yi (1994), whereas this is not seen in the approach of Abramowicz et al. (1995).

## 3. TOPOLOGY OF THERMAL EQUILIBRIA

The above equations were solved for various choices of the four disk parameters: the mass of the central black hole $M$, the mass accretion rate $\dot{M}$, the viscosity coefficient $\alpha$, and the distance from the central object $R$. It is convenient to discuss the results in the $\log(\dot{M}/\dot{M}_{\rm E})$ vs. $\log \Sigma$ plane, where $\dot{M}_{\rm E} = L_{\rm E}/c^2 = (4\pi c G M)/\kappa_{\rm es} c^2$ is the Eddington mass accretion rate. ($\kappa_{\rm es}$ is the opacity due to electron scattering.)

Figure 1 shows sequences of models corresponding to different choices of $\alpha$. The solutions refer to a fixed black hole mass $M = 10 M_\odot$, and to radii $R = 5 R_G$ (Fig. 1a) and $R = 30 R_G$ (Fig. 1b). At each $R$, there is a particular self-crossing sequence of models which corresponds to a critical value $\alpha_{\rm crit}$ of the viscosity parameter; $\alpha_{\rm crit}$ is a function of $R$, reaching a minimum value of $\approx 0.21$ at $R \sim 7 R_G$ and reaching unity at $R \sim 126 R_G$.

Figure 1 shows that the two $\alpha = \alpha_{\rm crit}$ curves divide the $\log(\dot{M}/\dot{M}_{\rm E}) - \log \Sigma$ plane into four distinct sections, dividing the disk models into four classes. We discuss below the properties of these four types of solutions as a function of three parameters, namely, the viscosity parameter $\alpha$ (relative to $\alpha_{\rm crit}$), the effective optical depth $\tau_{\rm eff}$, and the ratio of advective cooling to heating, $q = Q_{\rm adv}/Q_+$. In the following we use the term slim disk to refer to systems with $H/R \lesssim 1$ as well as those with $H/R \approx 1$. The latter are often described as thick disks, but this distinction is not important since the slim disk equations appear to describe even thick systems quite well (Narayan & Yi 1995a).

*Type I: Geometrically thin ($H/R \ll 1$) or slim ($H/R \lesssim 1$), optically thick ($\tau_{\rm eff} \gg 1$) disks with small viscosity ($\alpha < \alpha_{\rm crit}$).* The equilibrium sequences are shown by the S curves in solid lines at the right of Figs. 1a, b. Models on the lower and middle branches of the S are the usual Shakura-Sunyaev (1973) solutions. Models on the upper branch are dominated by radiation pressure, cooled by advection, and have $q \to 1$, $H/R \lesssim 1$. This branch was discovered by Abramowicz et al. (1988). The lower and upper branches are thermally and viscously stable, whereas the middle branch is both thermally and viscously unstable.

*Type II: Geometrically thin ($H/R \ll 1$) disks with small advective cooling ($q \ll 1$) and large viscosity ($\alpha > \alpha_{\rm crit}$).* These solutions are shown by the dashed lines in Figs. 1a, b. The equilibrium sequence consists of three branches. The two branches on the right are the same as the middle and lower branches of Type I (but for a higher value of $\alpha$) and have $\tau_{\rm eff} > 1$. The branch on the left corresponds to a gas pressure dominated, optically thin ($\tau_{\rm eff} < 1$), viscously stable, but thermally unstable, flow. Such sequences of models were calculated by Shapiro, Lightman & Eardley (1976) and more recently by Luo & Liang (1994).

*Type III: Geometrically slim ($H/R \lesssim 1$), optically thin ($\tau_{\rm eff} \ll 1$) disks with small viscosity ($\alpha < \alpha_{\rm crit}$).* These solutions are shown by the dotted lines in Figs. 1a, b. The equilibrium sequence consists of two branches. The branch on the right is similar to the viscously stable, but thermally unstable, branch of Type II. The branch on the left is a stable gas pressure supported, optically thin, advection-dominated ($q \to 1$) slim disk ($H/R \lesssim 1$) solution. This branch was found independently by Abramowicz et al (1995) and Narayan and Yi (1995b) (see also Narayan & Popham 1993).

*Type IV: Geometrically slim ($H/R \lesssim 1$) disks with large advection ($q \to 1$) and large viscosity ($\alpha > \alpha_{\rm crit}$).* These solutions, shown by the long-dashed lines in Figs. 1a, b, are fully advection-dominated and extend over the entire range of $\dot M/\dot M_{\rm E}$, switching from being optically thin ($\tau_{\rm eff} < 1$) at low $\dot M$ to optically thick at high $\dot M$. This sequence corresponds to some of the sub-Keplerian, advection-dominated solutions found by Narayan and Yi (1994, 1995a), but the fact that they extend without break in $\dot M$ is a new result.

In order to test whether the results are specific to the model of Abramowicz et al (1995), we have redone the calculations with the model of Narayan & Yi (1995b) briefly outlined towards the end of §2. Figure 1c shows the topology of the solutions at $R = 5R_G$ for $\alpha = 0.1$ and 1 with this model. We see exactly the same four types of solutions as in Figs. 1a, b. The main differences are that the curves move to lower values of $\dot M$, because of the more efficient cooling due to Comptonized synchrotron emission, while the inclusion of magnetic pressure causes minor changes in the geometrically thin disk branch. Nevertheless, the *qualitative* features of the topology are preserved intact, and even the value of $\alpha_{crit}$ is very little changed: 0.41 in Fig. 1c against 0.25 in Fig. 1a. This test shows that the classification of solutions described above is quite robust and is independent of the details of the particular dynamical or cooling model employed. In fact, even when the viscosity coefficient $\nu$ is taken to be proportional to gas pressure rather than to the total pressure, the topology continues to remain the same. All four types of solutions exist, but the value of $\dot M_{\rm crit}$ increases substantially.

## 4. DISCUSSION OF THE VARIOUS STABLE EQUILIBRIA

Apart from showing the different topologies, Fig. 1c also indicates the stability properties of the solutions. The line segments which are shown by dashed lines, and identified by the letter U, are thermally unstable and therefore not relevant for real flows. The remaining segments, shown by solid lines, are both thermally and viscously stable. Solutions identified by the letter C represent cooling-dominated flows. Solutions marked A which occur at higher accretion rates describe advection-dominated flows, where cooling is inefficient, either because the medium is optically thin or too optically thick.

For each $R$ and $\alpha$, the C solutions terminate at a maximum $\dot{M}$ above which they are thermally and viscously (Lightman & Eardley 1974) unstable. The solid lines in Fig. 2 show this maximum $\dot{M}$ as a function of $R$ for three choices of $\alpha$: 0.01, 0.1, 1. There are no stable cooling-dominated thin disk solutions above these lines.

The advection-dominated A solutions come in several varieties. For $\alpha < \alpha_{crit}$, these solutions split into two branches, those at small $\dot{M}$ which are advection-dominated because they are optically thin and unable to cool, and those at large $\dot{M}$ which are extremely optically thick and again unable to cool (Narayan & Yi 1994, 1995a). In between these two, there is a range of $\dot{M}$ where there is no advection-dominated solution allowed. The new and unexpected feature revealed by the present calculations is that, for $\alpha > \alpha_{crit}$, the A solutions span the entire range of accretion rates and are present for all values of $\dot{M}$. The critical parameter $\alpha_{crit}$ separates these two topologies and is a function of $R$. The resulting situation is indicated by the dashed and dotted lines in Fig. 2. Advection-dominated solutions are allowed only below and to the left of the dashed lines and above and to the left of the dotted lines.

A study of Fig. 2 reveals that four different possibilities arise in the $\dot{M}R$ plane. We identify the corresponding zones as 1–4. In zone 1, the only stable solution allowed is an advection-dominated flow, while in zone 2 only the cooling-dominated solution is allowed. In zone 3 both solutions are possible. Which of the two does nature pick? Perhaps the answer depends on the history of the accreting gas. Alternatively, a thermal instability in the upper layers of thin disks may cause such disks to switch to an advection-dominated form whenever the latter solution is available (Narayan & Yi 1995b).

In zone 4, there is no stable solution available. In analogy with models of dwarf-nova outbursts (see Cannizzo 1993 for a review) we suggest that in this region of the $\dot{M}R$ diagram, a thermal limit cycle is set up where the flow oscillates between two states (Abramowicz et al. 1988). Note, however, that Honma et al. (1991) found that the disk does not reach the upper, optically thick branch.

The exact locations of the various zones in the $\dot{M}R$ plane depend on the cooling model used, as seen from a comparison of Figs. 1a, c. In the presence of strong Comptonized synchrotron cooling, all the lines in Fig. 2 move down by up to 1.5 in log. One result is that systems with $\dot{M} \lesssim \dot{M}_{\rm E}$ are likely to sample different zones at different radii, so that we may expect quite interesting phenomena in these systems accompanied perhaps by complex spectral signatures. This may explain the variety of spectral states observed in X-ray binaries and active galactic nuclei.

Figure 2 shows that advection-dominated solutions are available over a surprisingly wide range of conditions. Indeed, if the advection-dominated branch is preferred over the cooling-dominated one in zone 3, as we speculate above, then it may well be that advection-dominated accretion is the most common form of accretion in black hole systems (especially at large $\alpha$), while standard thin disks may occur less frequently.

We thank Edison Liang for useful discussions and the referee, Paul Wiita, for helpful comments. This work was supported in part by NSF Grant AST 9148279.

# REFERENCES

Abramowicz, M. A., Chen, X., Kato, S., Lasota, J. P., & Regev, O. 1995, ApJ, 438, L37

Abramowicz, M. A., Czerny, B., Lasota, J. P., & Szuszkiewicz, E. 1988, ApJ, 332, 646

Artemova, J., Bisnovatyi-Kogan, G., Björnsson, G., & Novikov, I. 1995, preprint

Cannizzo, J.K. 1993, in Accretion Disks in Compact Stellar Systems, ed J. Craig Wheeler, (World Scientific: Singapore), p. 6

Chen, X., & Taam, R. E. 1993, ApJ, 412, 254

Honma, F., Matsumoto, R. & Kato, S. 1991 PASJ, 43 147

Lasota, J. P., & Pelat, D. 1991, A&A, 249, 574

Lightman, A. P., & Eardley, D. N. 1974, ApJ, 187, L1

Luo, C., & Liang, E. P. 1994, MNRAS, 266, 386

Narayan, R., & Popham, R. 1993, Nature, 362, 820

Narayan, R., & Yi, I. 1994, ApJ, 428, L13

———. 1995a, in press

———. 1995b, submitted

Paczyński, B., & Wiita, P. J. 1980, A&A, 88, 23

Pringle, J.E., Rees, M.J. & Pacholczyk, A. G. 1973, A&A, 29, 179

Shakura, N. I., & Sunyaev, R. A. 1973, A&A, 24, 337

Shapiro, S. L., Lightman, A. P., & Eardley, D. N. 1976, ApJ, 204, 187

Shaviv, G., & Wehrse, R. 1989, in Theory of Accretion Disks, eds W. Duschl et al., (Kluwer: Dordrecht) p. 419

Wandel, A., & Liang, E. P. 1991, ApJ, 380, 84

# FIGURE CAPTIONS

**Figure 1.**— (a) Thermal equilibria of accretion disks at $R/R_G = 5$. The two heavy solid lines correspond to solutions with $\alpha = \alpha_{crit} = 0.2446$. The solid, dotted, dashed, and long-dashed lines represent respectively the four types of equilibrium disk sequences discussed in §3. The calculations were done using the equations given in §2. (b) Same as (a) but for $R/R_G = 30$. (c) Same as (a) but calculated using the model of Narayan & Yi (1995b). Note the close similarity to (a), but with all curves shifted down by about 1.5 in the log, primarily due to the effect of strong Comptonized synchrotron cooling.

**Figure 2.**— Zones in the $\log(\dot{M}/\dot{M}_{\rm E}) - \log(R/R_G)$ plane corresponding to various combinations of stable equilibria, as discussed in §4.

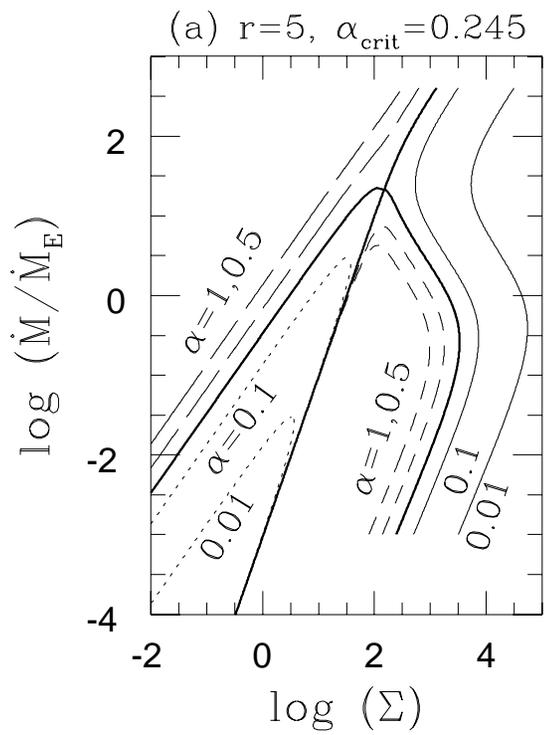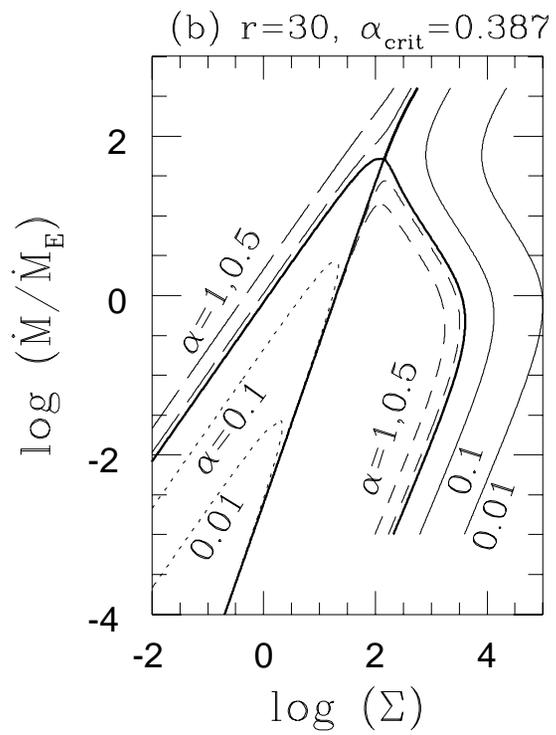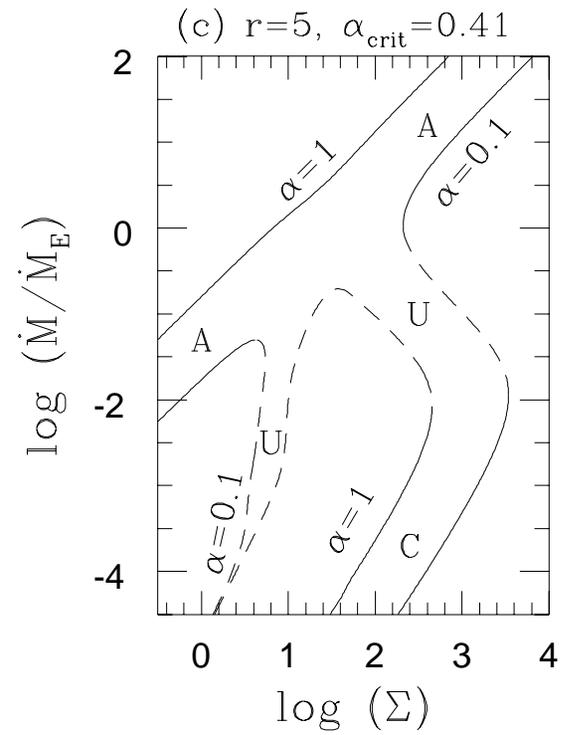

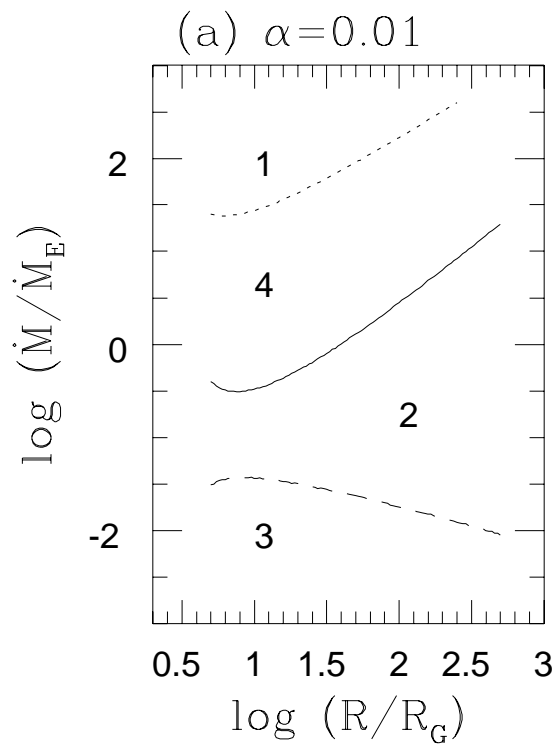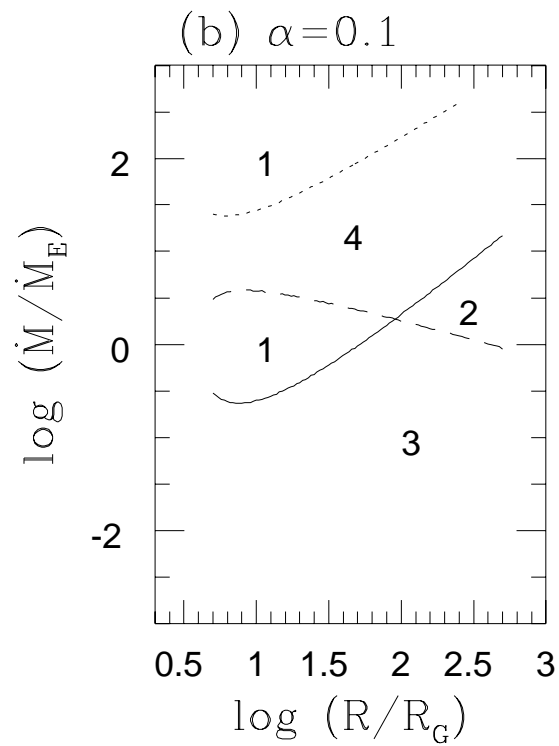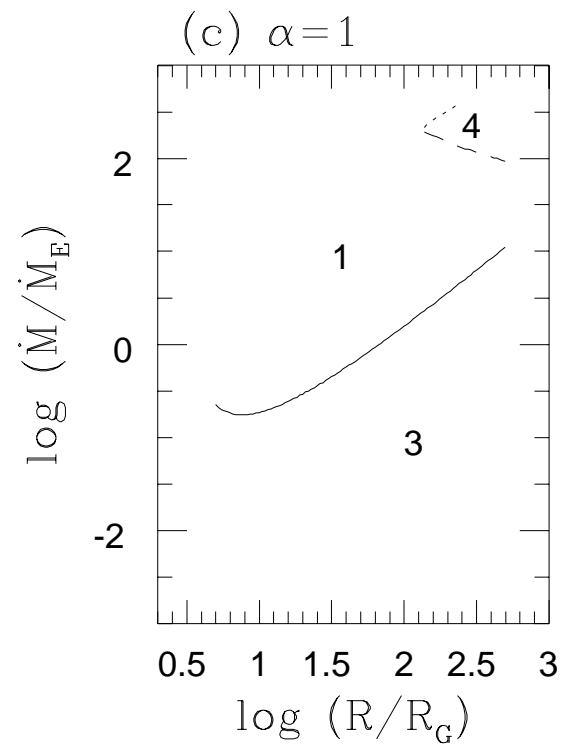